# Multimedia and Immersive Training Materials Influence Impressions of Learning But Not Learning Outcomes

Benjamin A. Clegg, Alex Karduna, Ethan Holen, Jason Garcia, Matthew G. Rhodes, & Francisco R. Ortega
Colorado State University
Fort Collins, Colorado
benjamin.clegg@colostate.edu, alexkarduna@gmail.com, eholen@rams.colostate.edu,
J.S.Garcia@colostate.edu, Matthew.Rhodes@ColoState.edu, F.Ortega@colostate.edu

## ABSTRACT

Although the use of technologies like multimedia and virtual reality (VR) in training offer the promise of improved learning, these richer and potentially more engaging materials do not consistently produce superior learning outcomes. Default approaches to such training may inadvertently mimic concepts like naïve realism in display design, and desirable difficulties in the science of learning – fostering an impression of greater learning dissociated from actual gains in memory.

This research examined the influence of format of instructions in learning to assemble items from components. Participants in two experiments were trained on the steps to assemble a series of bars, that resembled Meccano pieces, into eight different shapes. After training on pairs of shapes, participants rated the likelihood they would remember the shapes and then were administered a recognition test. Relative to viewing a static diagram, viewing videos of shapes being constructed in a VR environment (Experiment 1) or viewing within an immersive VR system (Experiment 2) elevated participants' assessments of their learning but without enhancing learning outcomes.

Overall, these findings illustrate how future workers might mistakenly come to believe that technologically advanced support improves learning and prefer instructional designs that integrate similarly complex cues into training.

## ABOUT THE AUTHORS

**Ben Clegg** is a Professor of Cognitive Psychology at Colorado State University. He earned his Ph.D. in Psychology from the University of Oregon. Dr. Clegg conducts research on Applied Cognitive Psychology and Human Factors.

**Alex Karduna** is an undergraduate student at Colorado State University interested in Augmented and Virtual Reality and Computer Science Theory.

**Ethan Holen** is a graduate student in computer science working on augmented and virtual reality research.

**Jason Garcia** is a Ph.D. student in the Math Department at Colorado State University.

**Matthew Rhodes** is a Professor of Cognitive Psychology at Colorado State University. His research focuses on memory, metacognition, cognitive aging, and evidence-based approaches to learning and training. He is also an author of a recent book on learning, *A Guide to Effective Studying and Learning: Practical Strategies from the Science of Learning* (2020; Oxford University Press).

**Francisco R. Ortega** is an Assistant Professor at Colorado State University and Director of the natural user interaction lab (NUILAB). Broadly speaking, his research has focused on multimodal and unimodal interaction (gesture-centric), which includes gesture recognition and elicitation (e.g., a form of participatory design). His main research area focuses on improving user interaction by (a) multimodal elicitation, and (b) developing interactive techniques. The primary domains for interaction include immersive analytics, assembly, and collaborative environments using augmented reality headsets.

# Multimedia and Immersive Training Materials Influence Impressions Of Learning But Not Learning Outcomes

**Benjamin A. Clegg, Alex Karduna, Ethan Holen, Jason Garcia, Matthew G Rhodes, & Francisco R. Ortega**
**Colorado State University**
**Fort Collins, Colorado**
**benjamin.clegg@colostate.edu, alexkarduna@gmail.com, eholen@rams.colostate.edu,**
**J.S.Garcia@colostate.edu, Matthew.Rhodes@ColoState.edu, F.Ortega@colostate.edu**

## INTRODUCTION

The addition of technologies like multimedia, virtual reality, augmented reality, and mixed reality to training contexts seems to offer the promise of improved learning. However, these advanced systems do not inherently produce superior outcomes. Despite the intuitive sense that delivering realistic, immersive, engaging, and enhanced information should increase training effectiveness, a recent meta-analysis revealed no systematic learning advantage from utilizing virtual and augmented reality in training (Kaplan et al., 2020).

The challenge of harnessing the right dimensions of technologies for training in optimal ways may parallel the earlier concept of *naïve realism* in display design (e.g., Smallman & St John, 2005). Naïve realism noted the tendency for designers and users to prefer realistic interfaces over simplified ones, putting a misplaced trust in their subjective sense of what seemed 'better'. Similarly, the domain of learning offers evidence of a disconnect between what individuals tend to prefer and believe will produce superior outcomes, and the practices and techniques that do optimize learning. For example, the notion of *desirable difficulties* (e.g., Bjork & Bjork, 2020) reflects the idea that requiring effort during learning better supports learning than more passive approaches. Although the science of learning indicates the presence of sufficient challenge is critical, the potential benefits from such approaches appear counterintuitive to most people (Rhodes, Cleary, DeLosh, 2020). False impressions about our own learning entails that many students and teachers routinely avoid studying techniques that require greater effort, despite this type of effort being linked to enhanced later memory for the information.

Dissociations between perceptions of learning and objective measures of learning outcomes have been documented in several domains. For instance, participants regard large relative to small words and loud relative to quiet presentations as fostering greater learning, even in the absence of a learning advantage (Rhodes & Castel, 2008; 2009). Within a multimedia learning context, participants may also overestimate learning benefits when shown static photos while learning concepts (Serra & Dunlosky, 2010), judging their comprehension of a text (Jaeger & Wiley, 2014; Linder et al., 2021) or learning foreign vocabulary terms (Carpenter & Olson, 2012). To be clear, multimedia support can frequently enhance learning (see e.g., Mayer & Mayer, 2005, for a review) but participants' perceptions of their own learning often fail to distinguish between instances in which such material serves to support learning (e.g., diagrams) versus serves a decorative purpose with no impacts on learning outcomes. Accordingly, judgments may reflect a multimedia heuristic, such that perceptions of learning uniformly increase when multimedia materials are included (e.g., Serra & Dunlsoky, 2010; Jaeger & Wiley, 2014; Linder et al., 2021).

A tendency for virtual and augmented environments to create misplaced subjective perceptions of superior learning, could potentially mislead both designers and learners. The present study thus sought to examine whether materials of the type generated by, and present within, virtual and augmented reality contexts offer the surface appearance of being richer, more engaging, and hence more memorable, while not actually actively employing dimensions in ways that drive greater learning. Accordingly, participants in two experiments were trained on the steps to assemble a series of bars, that resembled Meccano pieces, into eight different shapes. Experiment 1 compared performance after learning to assemble objects from pieces viewing step-by-step diagrams versus videos generated with a VR environment of the shapes being constructed. The research questions examined:
    (1) Whether the richer materials present in the video instructions would influence participants to provide higher metacognitive judgments of learning than the simpler static figures.
    (2) Whether the properties associated with those richer materials, in the absence of the explicit use of mechanism known to enhance memory, would produce any increase in learning.

**EXPERIMENT 1**

**METHODS**

The experiment followed a between-subjects design, with half of the participants viewing instructions on how to assemble objects in the form of static images instructions providing step by step sets of actions, and the other half following dynamic video instructions that reflected assembly captured in a VR environment.

**Participants**

Each participant gave informed consent prior to commencing the experiment. Data were collected from 130 people on Prolific, all of whom were located in the United States. Random assignment to condition produced 66 participants in the static diagram condition, and 64 participants in the video condition.

**Materials & Procedure**

All participants viewed materials on their own devices using a Qualtrics survey.

Participants were asked to learn the steps to assemble 8 different objects. Each object was constructed from 4 to 9 parts, and parts varied in size, having 3 to 8 possible positions to attach to other parts. To assemble the object the correct parts had to be placed in the correct relative position and orientation to each other using the correct attachment positions (an example object is shown in **Figure 1**).

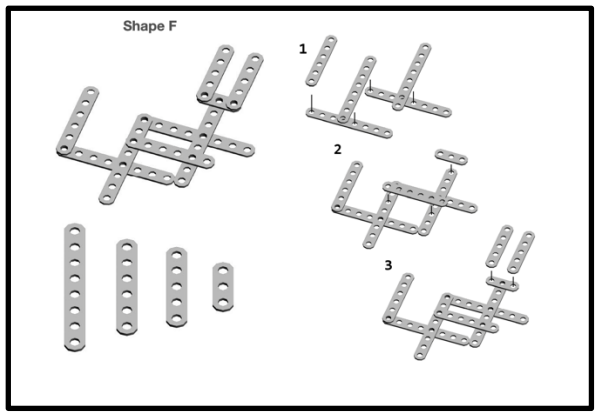

**Figure 1.** The static images assembly instruction.

Instructional materials to learn to build the shape varied between two different assigned conditions, with either video of the assembly or a static diagram of the steps involved. Static diagrams were developed by a graphic designer in Blender. These 3d model of pieces were combined in images showing assembly broken down into 3 steps, with each step connecting 1 to 4 parts to the object. The diagrams showed the final shape, the parts to be used and each of the 3 steps (see **Figure 1**).

Video assembly instructions were created from importing the Blender images used to create the static diagrams into the Unity game engine. In Unity we configured the scene so that it would run on an Oculus Quest (rather than the default of running on a monitor) using the Oculus Integration Plugin. The main feature used was the new camera, and this converted the scene to visuals in the Oculus. We then added the 3d models to the scene so that they could be moved, rotated, and scaled.

The videos were made by adding a script to each of the 3d models. This script moved pieces along a specified path so that when started, the object would move from the starting location to the location that it was supposed to be in the build. Once it reached its location it then triggered the next object to move to its target location in the build, and so on. Accordingly, in the video, objects moved from their start location to their target location in the expected order. Within the simulated scene, items appeared to be automatically selected from a menu of parts and then moved smoothly through space one at a time to their correct location on the object. The attachment position on the current part and the target attachment position on the object were highlighted using a green dot during motion (see **Figure 2**).

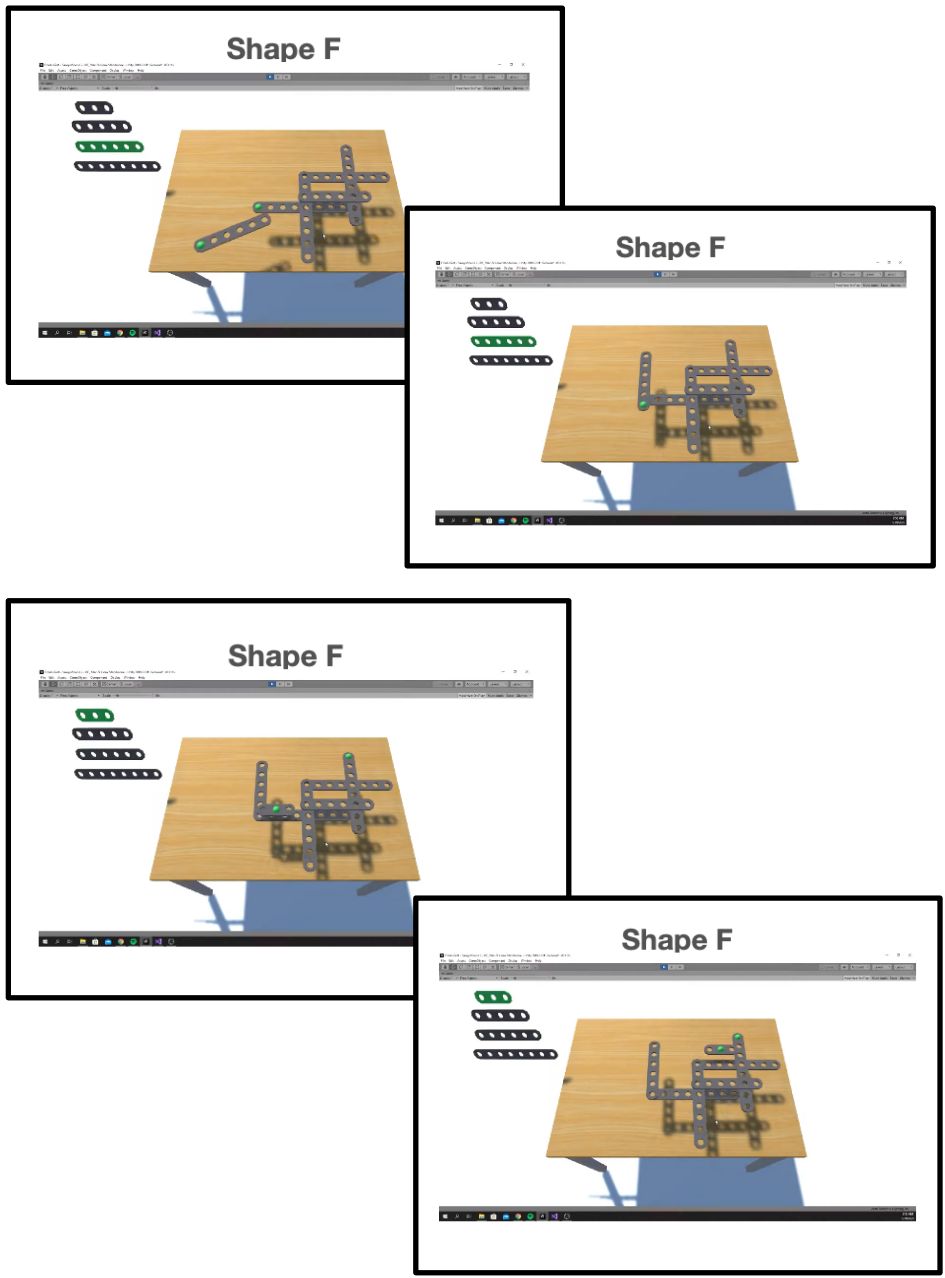

**Figure 2.** Screenshots from the video assembly instructions for the same object shown in Figure 1.

In studying how to assemble the objects, participants were warned they would have a limited time to learn each object. Participants were then shown a set of instructions the format of which varied with their assigned condition. Assembly instructions were shown for a fixed time period with a countdown in seconds to indicate the study time remaining. Time for each shape was matched to the time for the video condition of the instructions (with durations of approximately 30 seconds) and the page automatically advanced when time expired.

Training on the eight assembly tasks was divided into 4 blocks, with two different objects studied per block. Block order was randomized across participants. After studying the assembly of each object in the block, users were then presented with a question to assess their judgment of the learning that had occurred. Within each block after answering this metacognitive question for the first object, participants were then shown the instructions for the second assembly task. After the time allotted, they again indicated their perception of the learning for the second assembly task. Following the presentation of assembly instructions for two different objects, participants were presented six recognition test questions to index their learning.

**Measures**

Two main dependent variables were collected: (1) The participant's rating of their own learning following the instructions; and (2) the participant's score on a set of recognition tests.

To assess judgments of learning, after studying how to build objects, participants gave a response from 1 to 100 using a slider:
**What is this chance you will correctly remember this when you are tested later?**
**(Range 0%--No chance at all     100%--Entirely Certain I will Correctly Remember This)**

A three alternative forced choice test was used to assess knowledge of the trained shapes. The quiz for each object consists of two sets of three questions. One set probed an action required for the middle step in the construction (one of the pieces placed on step 2 of 3 in the static diagram), and the other set a late step in the construction of the object (step 3 of 3 in the static diagram). Participants were shown an image of the task at an uncompleted stage of the assembly. They were then asked about the next step in the assembly. The three questions were: (I) **Which bar comes next?** (II) **Which holes are used to attach the bar?** (III) **Where does the bar go?**

For each question, the respondents were offered three visual choices to select from. The order of the choices shown to the participant was randomized. A sample question can be seen below in **Figure 3**.

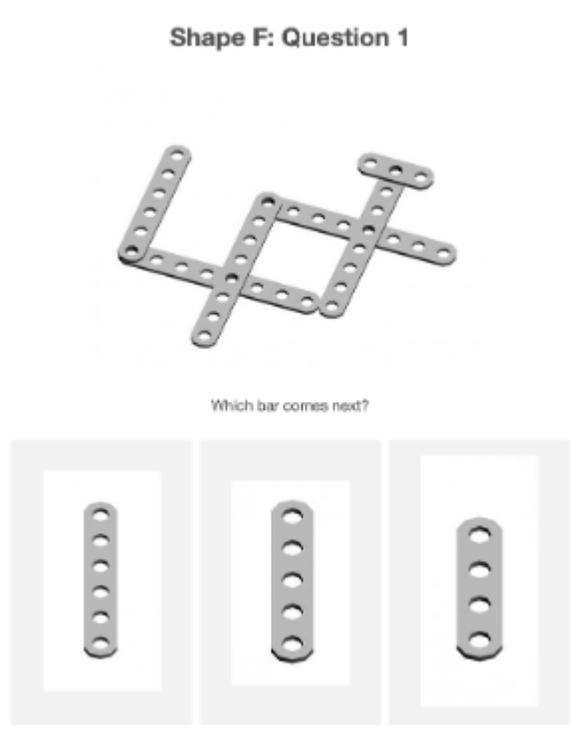

**Figure 3:** Example three alternative forced choice test of which piece came next. These tests were used to assess recognition of required assembly steps.

**RESULTS**

A comparison of the average subjective ratings of learning following study of pairs of items (see **Figure 4**) showed that participants provided significantly greater ratings of learning from the video assembly instructions (Mean rating: 54.2, *SD*: 16.7) than the static diagrams (*M*: 44.0, *SD*:20.6), *t*(128) = 3.11, *p* < .005, *d* = 0.54. In contrast to the greater impression of learning, scores on the memory test (see **Figure 5**) showed no difference between the video assembly instructions (mean percent correct: 71.8%, *SD*: 12.1) and the static diagrams (*M*: 72.8%, *SD*:15.3), *t*(128) = 0.42, *p* > .05, *d* = 0.07. To assess whether the lack of difference might have been an artifact of learning metric, we also calculated performance on the assembly task looking at the percentage of objects for which all six questions were answered correctly. This again showed no impact of the type of instruction (*t*(128) = 0.63, *p* > .05, *d* = 0.11), with the video assembly instructions (mean shapes completely correct: 33.1%, *SD*: 20.0) comparable to the static diagrams (*M*: 30.7%, *SD*: 23.8).

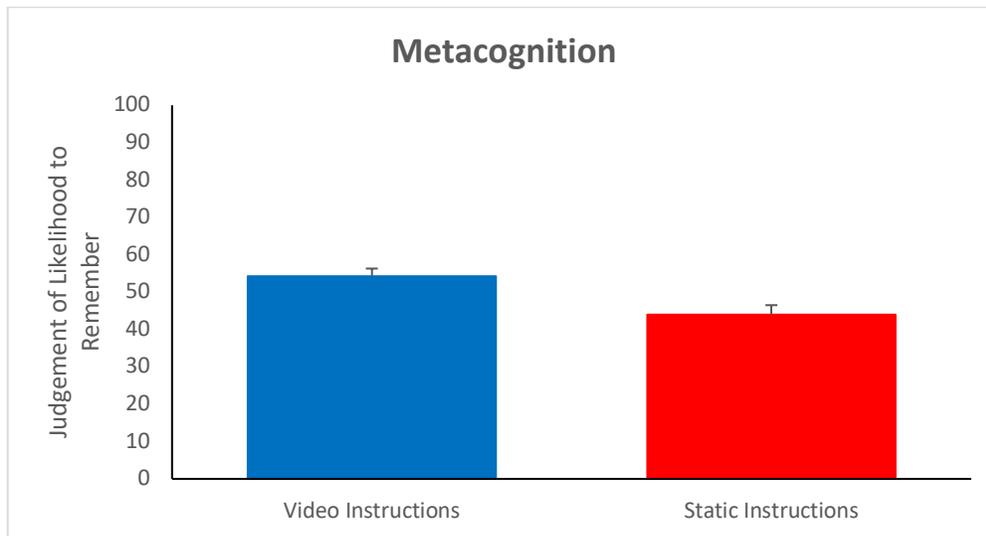

**Figure 4.** Experiment 1: judgments of perceived learning after training on object assembly by condition.

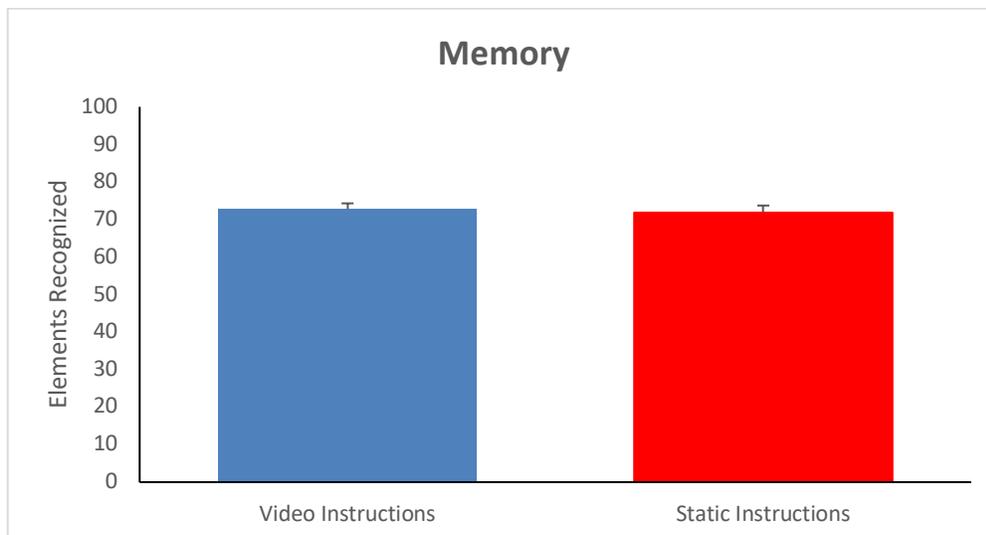

**Figure 5.** Experiment 1: percentage of questions correct on a recognition test for the assembly process steps by condition.

An additional set of analyses explored performance on those tests from the earlier steps of construction versus those later in the assembly process. Here there is a contrast between the lower memory load from fewer pieces at an earlier stage versus the greater contextual cues from the existing pieces at the later stage. A mixed 2 (instruction type) x 2 (middle or late assembly step) analysis of variance (ANOVA) showed that later steps (*M*: 78.2%, *SD*: 15.5) were better identified than earlier steps (*M*: 66.4%, *SD*: 15.1), $F(1,128) = 102.05$, $p < .0005$, $\eta^2_p = 0.44$. There was again no effect of instruction type ($F(1,128) < 1$, $\eta^2_p = 0.01$), nor an instruction by step interaction ($F(1,128) < 1$, $\eta^2_p = 0.01$).

**DISCUSSION**

The results of Experiment 1 showed that materials from the video did not change learning outcomes compared to more standard, drawing-based instructions. However, the richer materials in the videos did elevate confidence in learning, suggesting that the types of features found in advanced technologies may promote a false sense of greater learning. Thus, both users being trained within these systems and those designing the training may need to be appropriately wary that the introducing materials that appear more compelling can risk simply becoming an expensive and time-consuming façade that serves no training purpose.

**EXPERIMENT 2**

Experiment 2 sought to further increase the richness of the appearance of the materials used to study the object assembly by adding a passive viewing of the training information within an immersive VR system condition. The research was intended to explore whether even greater richness in the training materials further inflated perceptions of learning or whether the presence of even more engaging aspects might trigger benefits in training outcomes.

**METHODS**

Except where noted, the methods for Experiment 2 mirrored those of Experiment 1. As in Experiment 1, Experiment 2 followed a between-subjects design. Half of the participants followed video instructions, similar to those used in Experiment 1, whereas the other half were given immersive VR instructions.

The immersive VR instructions were also created using Unity as with the videos used in Experiment 1. To retain consistency between the experiments, the same scene was used for both. The only difference was that a VR controller was added for the virtual scenes. Other minor changes were made to the positioning of the bars so as to solve perspective shift issues as the participants would now be viewing the shapes from different angles and not exclusively from the top-down as in the video.

The immersive VR instructions were presented to each participant in the Oculus Quest VR headset. Participants were asked to put on the headset each time a video was played and to let the experimenter know when they could clearly view the table. The animation which was recorded in the video was then played. After the animation was completed, each participant was then given 5 seconds to observe the object. The participant was then asked to remove the headset and answer the questions related to the specific shape.

**Participants**

Participants were recruited via email and flyers on a university campus. A total of 34 participants were randomly assigned to the two forms of instructions, learning to assemble objects from either video of the steps ("Video Instructions") or watching steps in an immersive, but passive VR environment ("VR Passive") with 17 people in each condition.

**RESULTS**

As in Experiment 1, we again explored both perceived learning and actual memory for the information studied.

A comparison of the average subjective ratings of learning following study of pairs of items showed no significant difference between the video assembly instructions (Mean rating: 58.9, *SD*: 15.7) relative to 3D viewing (*M*: 60.3, *SD*:15.8), t(32) = 0.25, $p > .05$, $d = 0.09$. In addition, scores on the memory test did not differ between the video assembly instructions (mean percent correct: 75.1%, *SD*: 7.5) and the immersive viewing of assembly (*M*: 76.8%, *SD*: 8.8), t(32) = 0.61, *p* > .05, *d* = 0.21.

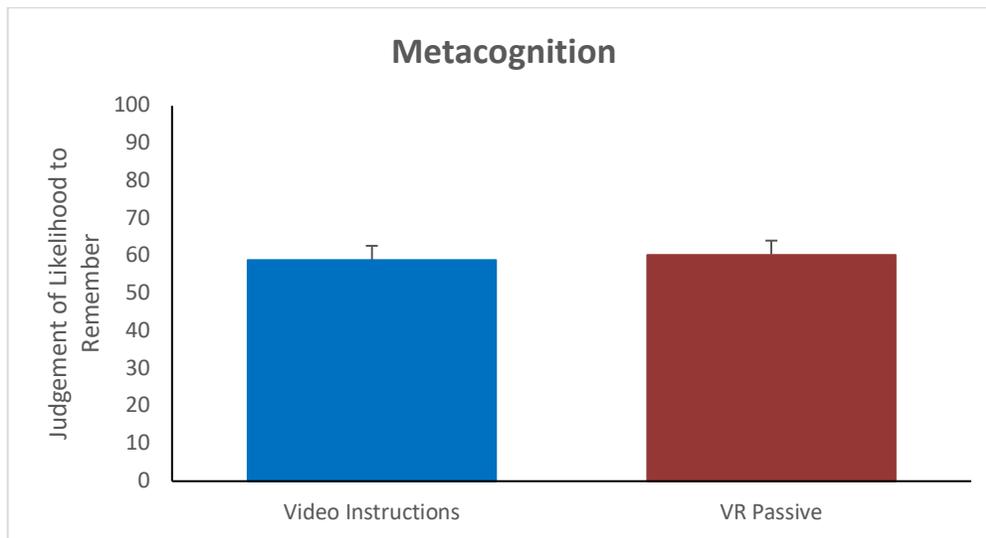

**Figure 6.** Experiment 2: judgments of perceived learning after training on object assembly by condition.

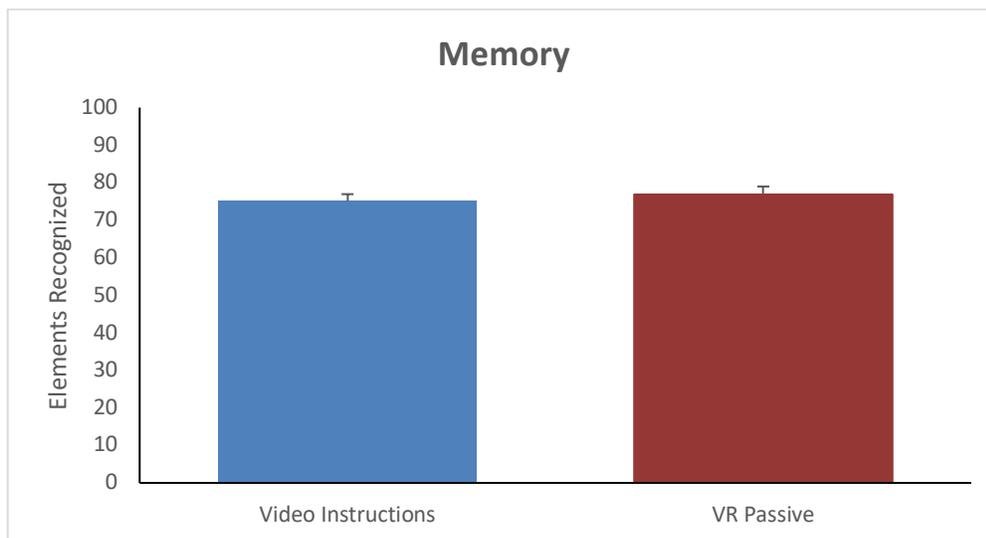

**Figure 7.** Experiment 2: percentage of questions correct on a recognition test for the assembly process steps by condition.

An analysis combining the two experiments comparing the groups showed a significant difference in the ratings of learning (F(3,160) = 6.46, p < .001, $\eta^2_p$ = 0.11) and, most importantly, both Experiment 2 groups with richer materials also differed in their judgments of the learning that had occurred following study compared to the previous static image condition from Experiment 1 (Exp. 1 static versus Exp. 2 video: t(81) = 2.78, p < .01, d = 0.376; Exp. 1 static versus Exp 2 VR passive: t(81) = 3.03, p < .005, d = 0.83). However, an analysis of memory for the information showed no significant differences in the learning between any of the four groups (F(3,160) < 1, $\eta^2_p$ = 0.02), including for the passive VR condition compared to the previous static image condition from Experiment 1 (Exp. 1 static versus Exp 2 VR passive: t(81) = 1.30, p > .05, d = 0.35)

**DISCUSSION**

The results of Experiment 2 suggest the richness of materials do not necessarily directly translate into greater self-perceptions of learning. That is, presenting training within a VR environment produced no differences in the ratings of learning that occurred compared to the more basic video viewing condition. In addition, as in Experiment 1, there were no changes to the actual learning outcomes from richer materials in this experiment.

Although no differences in learning or judgments of learning emerged, a cross-experiment comparison showed that metacognitive judgments were still elevated compared to static diagrams. Even richer cues in the VR viewing condition did not further accentuate judgments of learning compared to the video. This highlights factors that may and may not influence our sense of learning. Some aspects of the animated, detailed, three dimensional, and continuous movement features in the videos were sufficient to create a sense of fluency about studying that inflated the self perception of the learning that was occurring. However, including additional real-world fidelity, such as an immersive, stereoscopic presentation environment did not heighten the sense of improved learning any more than the video renderings had created.

**GENERAL DISCUSSION**

In two experiments, we examined the influence of rich instructional materials, including instruction filmed and viewed in a VR environment, on perceptions of learning and learning outcomes. Overall, our findings suggest that such multimedia instruction may engender an illusion of mastery: Relative to viewing static images, participants who viewed a video of assembly recording from VR or viewed within VR showed elevated perceptions of learning. In contrast to these perceptions, learning outcomes did not differ across any of the instructional formats investigated. Overconfidence in learning is reflected in the fact that participants rated the chances they would remember how to assemble objects as greater than 2/3 of the time on average, whereas their actual ability showed that only about 1/3 of the time were they able to correctly completely identify all the critical information: the pieces, the location, and the position for attachment. Moreover, the probes only examined a small subset of steps; it is possible the participants' actual accuracy at constructing the objects would have been even lower, if the gaps in their knowledge were on those other untested steps.

Our data highlight another instance of assessments of learning that over-weight the value of multimedia stimuli (e.g., Jaeger & Wiley, 2014). Indeed, these findings illustrate how future workers might mistakenly come to believe that technologically advanced support is enhancing their learning, even if it does not. These findings also set an agenda for future research to better understand when such immersive instruction benefits learning and when assessments of learning are at odds with learning outcomes.

It remains a question for future research as to what dimensions present in the environment for videos viewed on screens or with an immersive presentation cue a higher self perception of learning than static images (for example, dynamic motion or higher fidelity images of the objects). The current study would suggest that these misleading features were similarly present in all video conditions; interestingly, the immersive, stereoscopic view added no additional cues to drive even more inflated confidence in learning. The current findings also raise the importance of investigating whether active participation in training in virtual and augmented environments also serves to heighten judgments of the learning that is occurring, and whether, as in the current case, those too are dissociated from actual improvements in training outcomes.

Finally, the current findings offer an important cautionary note for those designing training using technologies that produce rich environments. Those that design such forms of instruction might be tempted by those same cues that mislead learners, and find themselves leaning towards developing complex forms of training that may not be producing superior training outcomes to far less sophisticated instructional materials.

**ACKNOWLEDGEMENTS**

This research was supported by the National Science Foundation under award number 1928502, program officer Dr. Betty Tuller.